\documentclass[aps,prd,twocolumn,showpacs,amsmath,nofootinbib]{revtex4-1}

\usepackage{color}
\newcommand{\beqa}{\begin{eqnarray}}
\newcommand{\eeqa}{\end{eqnarray}}
\usepackage{float}
\usepackage{epsfig}
\usepackage{graphicx}
\usepackage{latexsym}
\usepackage{amsmath}
\usepackage{mathrsfs}
\usepackage{dcolumn}% Align table columns on decimal point
\usepackage{bm}%

\begin{document}

%\title{Constraints on small-scale matter power spectrum with the global 21-cm signal}
\title{Influence of the deviation of the matter power spectrum at small scales on the global 21-cm signal at cosmic dawn}

\author{Yupeng Yang$^1$}\email{ypyang@aliyun.com}
\author{Xiujuan Li$^2$}\email{lxj@qfnu.edu.cn}
\author{Gang Li$^1$}\email{gli@qfnu.edu.cn}
\affiliation{$^1$School of Physics and Physical Engineering, Qufu Normal University, Qufu, Shandong, 273165, China \\
             $^2$School of Cyber Science and Engineering, Qufu Normal University, Qufu, Shandong, 273165, China}

\begin{abstract}
The matter power spectrum has been strongly constrained by astronomical measurements at large scales, 
but only weakly at small scales. 
Compared with the standard scenario, the deviation of the matter power spectrum at small scales 
has influence on the cosmological structure formation, e.g., the comoving number 
density of dark matter halos. The thermal history of the intergalactic medium (IGM) can be changed 
if dark matter is made of weakly interacting massive particles and can annihilate into 
standard model particles. The changes of the evolution of IGM could leave imprints on the relevant 
astronomical observations. 
Taking into account the dark matter annihilation, we investigate the impact 
of the deviation of matter power spectrum at small scales on the global 21-cm signal. 
In view of the measurements of the global 21-cm signal by the EDGES experiment, 
we explore the allowed parameter space of $m_s$, which describes the degree of deviation, by requiring 
the differential brightness temperature of the global 21-cm signal $\delta T_{21} \le -50~\rm mK$ at redshift $z=17$. 

\end{abstract}

%\keywords{}
\maketitle

\section{introduction} 
The standard inflation model has predicted that the primordial power spectrum is in a scale invariant form of 
$\mathcal{P}(k)\sim k^{n_{s}-1}$~\cite{model_1,model_2,model_3,planck-2018,Leach:2005av}. 
At large scales, $10^{-4}\lesssim k\lesssim 1~\rm Mpc^{-1}$, 
primordial power spectrum has been well 
constrained by astronomical measurements, e.g., cosmic microwave background (CMB), 
large-scale structure and Lyman-$\alpha$ 
forest~\cite{cmb_2,lyman,large}. 
At small scales, $k\gtrsim 1~\rm Mpc^{-1}$, the constraints are from the studies of, e.g., primordial black holes, 
ultracompact minihalos, galaxy luminosity functions and Silk damping effects~\cite{Josan:2009,Dalianis:2018ymb,mnras,prd-2020,scott_2015,Bringmann_1,
prd-2017,fangdali,Jeong:2014gna,Yoshiura:2020soa,prl_2}. 
The primordial power spectrum results in a matter power spectrum $P_{m}(k)\sim k^{n_s}$. The astronomical measurements such as CMB have been used to reconstruct  
the matter power spectrum at large scales $10^{-3}\lesssim k\lesssim 0.19~\rm Mpc^{-1}$~\cite{Hlozek:2011pc}. 
Large-scale 21-cm measurements could be used to probe the matter power spectrum at small scales~\cite{Munoz:2019hjh}. 
Any deviation of $P_{m}(k)$ at small scales 
can result in the changes of the cosmological structure formation such as the comoving number density of 
dark matter halos, while no conflict with existing astronomical measurements~\cite{Villanueva-Domingo:2021cgh,
Tashiro:2012qe,03480,Yoshiura:2018zts,Libanore:2022ntl}.

The existence of dark matter (DM) has been confirmed by many different astronomical observations. 
However, the nature of DM still keeps unknown. Different DM models have been proposed and one of the mostly 
studied is weakly interacting massive particles (WIMPs)~\cite{Bertone:2004pz,Jungman:1995df}. 
The relevant theory proposes that WIMPs 
can annihilate into standard model particles such as electrons, positrons and photons. 
These particles have interactions with that existing in the Universe, resulting in the changes 
of the thermal history of intergalactic medium (IGM)~\cite{binyue,xlc_decay,lz_anni,energy_function,DM_2015,
Slatyer:2015kla,Cheung:2018vww,Kovetz:2018zan,Berlin:2018sjs,DAmico:2018sxd,prd-edges,PhysRevD.103.123002,
chi_1,chi_2,Cumberbatch:2008rh,epjplus-2}. These changes 
could leave imprints on different astronomical observations such as the CMB and global 21-cm signal. 
Furthermore, the properties of DM can be investigated by the relevant astronomical measurements; 
see, e.g., Refs~\cite{DM_2015,
lz_anni,energy_function,Kovetz:2018zan,Berlin:2018sjs,DAmico:2018sxd,Valdes:2007cu,Vald_s_2012}. 

As mentioned above, the deviation of the matter power spectrum at small scales can lead to the changes 
of the comoving number density of DM halos. Taking into account the DM annihilation, 
it is expected that the deviation can lead to the different thermal history of the IGM 
and astronomical observations compared with the standard scenario. 
The authors of \cite{03480} have investigated these effects on the CMB observations. In this work, following the methods 
in \cite{03480} we will study the impact of the deviation of matter power spectrum at small scales 
on the global 21-cm signal in the cosmic dawn.

As an important way to study the early universe, the detection of the global 21-cm signal 
is very challenging~\cite{Pritchard:2011xb,Furlanetto:2006jb}. Recently, the Experiment to Detect the Global Epoch of 
Reionization Signature (EDGES) reported their results of the global 21-cm signal~\cite{edges-nature}. 
They found an absorption signal centered at redshift 
$z\sim 17$ about twice as large as expected~\cite{cohen-mnras,Furlanetto:2006jb,Xu:2021zkf}. Note that the results of the EDGES experiment 
are still controversial and require further verification~\cite{Bradley:2018eev,Hills:2018vyr,Singh:2019gsv,Singh:2017cnp}. 
On the other hand, 
the global 21-cm signal can be used to investigate 
the properties of DM; see, e.g, Refs~\cite{prd-edges,PhysRevD.103.123002,Fialkov:2018xre,Cumberbatch:2008rh,
Fraser:2018acy,Hektor:2018lec,Burns:2019zia,yinzhema,Kovetz:2018zan,Hiroshima:2021bxn,
Halder:2021rbq,Kovetz:2018zes,Bhatt:2019qbq,Berlin:2018sjs,Barkana:2018cct,Jia:2019yhr,DAmico:2018sxd,
PhysRevD.103.123002,Saha:2021pqf}. 
In this paper, taking into account the DM annihilation and by requiring 
the differential brightness temperature of the global 21-cm signal, e.g., $\delta T_{21} \lesssim -50~\rm mK$ at redshift $z=17$, 
we explore the parameter space of $m_s$, which characterizes the deviation of the matter power spectrum at small scales. 
Here we have not included the heating effects from 
astrophysical processes performed in the standard scenario~\cite{Minoda:2021vyw,Yoshiura:2019zxq,Xu:2021zkf,Cen:2016htm}.

This paper is organized as follows. In Sec. II we present the basically related components of the matter power spectrum considered here, 
and the basic equations for the evolution of the IGM including DM annihilation. 
In Sec. III, we investigate the impact of the deviation of matter power spectrum at small scales on the global 21-cm signal, 
and then explore the allowed space of the related parameter. The conclusions are given in Sec. IV. 
Throughout the paper we will use the cosmological parameters from Planck-2018 results~\cite{planck-2018}.

%%%%%%%%%%%%%%%%%%%%%%%%%%%%%%%%%%%%%%%%%%%%%%%

\section{The matter power spectrum at small scales and the evolution of IGM}

In the standard scenario, the matter power spectrum resulted from the primordial power spectrum is in a form of $P_{m}(k)\sim k^{n_s}$. 
Many other inflation models have been proposed and suggested that the primordial power spectrum could be deviated 
at small scales while being consistent with existing astronomical measurements 
at large scales. 
For the most inflation models, the deviation of $\mathcal{P}(k)$ is suggested in a form of power law growth 
at small scales; see, e.g., Refs~\cite{Byrnes:2018txb,Raveendran:2022dtb,Heydari:2021qsr,Carrilho:2019oqg,
Cole:2019zhu,Mishra:2019pzq,Yi:2020cut,Gao:2018pvq,Balaji:2022zur}. In view of these factors, following Ref.~\cite{03480}, 
we take the parametrized form of the matter power spectrum as follows

\beqa
P_{m}(k)=
\begin{cases}
A_{m}k^{n_s}~~~~~~~~~~~~~~~~k\le k_{p} \\
A_{m}k_{p}^{n_{s}}\left(\frac{k}{k_{p}}\right)^{m_s}~~~~~k> k_{p}
\end{cases}
\label{eq:pm}
\eeqa 
where the pivot scale $k_{p}\gtrsim 10~\rm Mpc^{-1}$ in order to be consistent with the available astronomical observations. 
The matter power spectrum at redshift $z$ can be written as 

\beqa
P_{m}(z,k) = P_{m}(k)T^{2}(k)\frac{D^{2}(z)}{D^2(0)},
\eeqa
where $D(z)$ is the growth factor~\cite{1992ARAA,Green:2005fa}, and $T(k)$ is transfer function~\cite{1986ApJ}. 
$A_{m}$ is a constant normalized as $\sigma_{8}=0.8$, where $\sigma_8$ is the root mean square mass 
fluctuation in a sphere of radius $8 h^{-1}~\rm Mpc$. The mass variance $\sigma^{2}(z)$ is written as follows

\beqa
\sigma^{2}(z,M) = \int \frac{dk}{k}\frac{k^{3}P_{m}(z,k)}{2\pi^{2}}W^{2}(kR),
\eeqa
where $W(x)$ is the window function and we use the form as

\beqa
W(x) = \frac{3\left({\rm sin}x-x{\rm cos}x\right)}{x^{3}}.
\eeqa

Since the changes of $P_{m}(k)$ investigated here are mainly on small scales, nonlinear effects 
are very important. There are several ways to deal with nonlinear effects. For the purpose of this work, one way is using the Zeldovich approximation or 
Lagrangian perturbation theory~\cite{White:2014gfa,non-linear,Porto:2013qua}. 
Another way is using the Press-Schechter(PS) theory~\cite{1974ApJ...187..425P}, which has been proved to be valid and wildly used 
in literature see, e.g., Refs.~\cite{10.1093/mnras/262.4.1023,Percival:2001nv,DODELSON2003261}. 
The evolution of nonlinear effects will result in the collapse of the regions with large density perturbation. Although on small scales 
the mass variance $\sigma^{2}$ calculated using the linear power spectrum is different from that of nonlinear power spectrum, PS theory 
shows that the collapsed fraction can be obtained using the linear power spectrum. In this work, we will use PS theory 
to deal with nonlinear effects on small scales. On the other hand, since we mainly focused on the effects of dark matter annihilation within 
dark matter halos, there is another method of calculating the 'boost factor'(BF) to deal with nonlinear effects. The BF can be obtained 
by directly integrating the nonlinear matter power spectrum for investigated scales at different redshifts~\cite{Serpico:2011in,Sefusatti:2014vha,Hiroshima:2021bxn}. Essentially, this method is the same as the PS theory.

\begin{figure}[htp]
\centering
\includegraphics[width=0.85\linewidth]{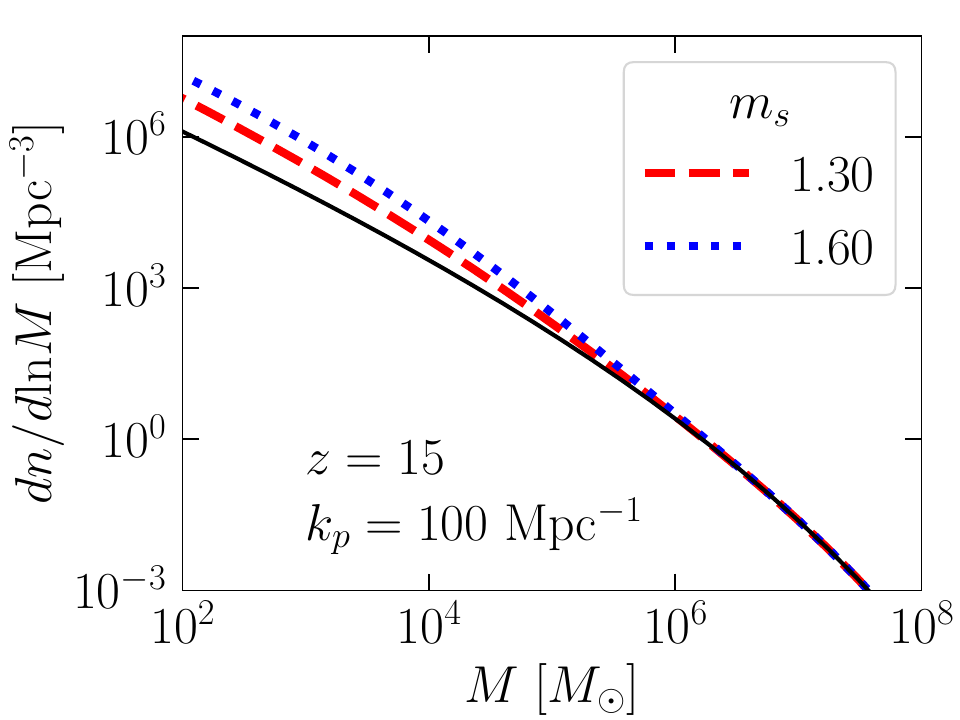}
\caption{Comoving number density of dark matter halos for different values of 
$m_{s}=1.30~{\rm and}~1.60$ at redshift $z=15$ for the pivot scale 
$k_{p} = 100~\rm Mpc^{-1}$. We also plot the standard scenario for comparison ($m_{s}=n_s=0.96$, thin solid black line). }
\label{fig:dndm}
\end{figure}

For the Press-Schecter mass function, the comoving number density of DM halos is in a form of~\cite{ps} 

\beqa
\frac{dn(z,M)}{dM}=\sqrt{\frac{2}{\pi}}\frac{\rho_{0}}{M}
\frac{\delta_{c}}{\sigma^{2}}\frac{d\sigma}{dM} \
{\rm exp}\left(-\frac{\delta_{c}^2}{2\sigma^{2}}\right),
\eeqa
where $\delta_{c}=1.686$ is the threshold for spherical collapse. 
In Fig.~\ref{fig:dndm}, we plot the comoving number density of DM halos for different values of 
$m_{s}=1.30~{\rm and}~1.60$ at redshift $z=15$ for pivot scale 
$k_{p} = 100~\rm Mpc^{-1}$. For comparison, we also plot the standard scenario with $m_{s}=n_s=0.96$~\cite{planck-2018}. 
From this plot, it can be seen that the deviation of the matter power spectrum at small scales 
results in an increase of the comoving number density of DM halos with small masses. 
Since we have set the pivot scale $k_{p}=100~\rm Mpc^{-1}$, 
compared with the standard scenario, the significant difference appears 
for the masses of $M \lesssim 6\times 10^{5}~M_{\odot}$. 

%%%%%%%%%%%%%%%%%%%%%%%%%%%%%%%%%%Figure 1  begin%%%%%%%%%%%%%%%%%%%%%%%%%%%%%%%%%%%%%%

%%%%%%%%%%%%%%%%%%%%%%%%%%%%%%%%%Figure 1 end %%%%%%%%%%%%%%%%%%%%%%%%%%%%%%%%%%%%%%%%%

Taking into account the DM annihilation, the energy release rate per unit volume 
can be written as~\cite{03480,xlc_decay,lz_anni,energy_function,DAmico:2018sxd,fermi-3} 

\beqa
\frac{dE}{dVdt}{\bigg |}_{\rm DM} = (1+z)^{3}\frac{\left<\sigma v\right>}{m_{\chi}}\int dM\frac{dn}{dM} 
\int 4\pi r^{2}\rho^{2}_{\chi}(r) dr,
\label{eq:de_dvdt}
\eeqa
where $\left<\sigma v\right>$ is the thermally averaged cross section of DM annihilation, 
and $m_{\chi}$ is the mass of DM particle. $\rho_{\chi}(r)$ is the density profile 
of DM halos and we adopt the Navarro-Frenk-White model for our calculations~\cite{NFW}.

%\subsection{The evolution of IGM including accreting PBHs}

The energy released from DM annihilation can inject into the Universe resulting in the changes 
of the thermal history of IGM. The evolutions of the ionization fraction $x_e$ and 
kinetic temperature $T_k$ of the IGM are governed by the following equations~\cite{mnras,DM_2015,xlc_decay,lz_decay}: 

\beqa
(1+z)\frac{dx_{e}}{dz}=\frac{1}{H(z)}\left[R_{s}(z)-I_{s}(z)-I_{\rm DM}(z)\right],
\label{eq:xe}
\eeqa

\beqa
(1+z)\frac{dT_{k}}{dz}=\frac{8\sigma_{T}a_{R}T^{4}_{\rm CMB}}{3m_{e}cH(z)}\frac{x_{e}(T_{k}-T_{\rm CMB})}{1+f_{\rm He}+x_{e}} \nonumber \\
-\frac{2}{3k_{B}H(z)}\frac{K_{\rm DM}}{1+f_{\rm He}+x_{e}}+2T_{k}, 
\label{eq:tk}
\eeqa
where $R_{s}(z)$ is the standard recombination rate, $I_{s}(z)$ is the ionization rate by standard sources. 
$I_{\rm DM}$ and $K_{\rm DM}$ are the ionization and heating rate by DM annihilation, 
which can be written as follows~\cite{lz_decay,xlc_decay,mnras,yinzhema,DM_2015}, 

\beqa
I_{\rm DM} = f_{i}(z)\frac{1}{n_b}\frac{1}{E_0}\frac{{\rm d}E}{{\rm d}V{\rm d}t}\bigg|_{\rm DM} 
\label{eq:I}
\eeqa
\beqa
K_{\rm DM} = f_{h}(z)\frac{1}{n_b}\frac{{\rm d}E}{{\rm d}V{\rm d}t}\bigg|_{\rm DM} 
\label{eq:K}
\eeqa
where $n_b$ stands for the baryon number density and $E_0= 13.6~\rm eV$. 
$f(z)$ is the fraction of the energy released from DM annihilation injected into the IGM for 
ionization and heating, respectively. Here we have used the public code ExoCLASS~\cite{exoclass}, 
a branch of the public code CLASS~\cite{class}, to calculate $f(z)$ numerically.

\section{The impact of the deviation on the global 21-cm signal and corresponding constraints}

The quantity associated with the observations describing the global 21-cm signal is 
the differential brightness temperature $\delta T_{21}$. Relative to the CMB background, 
$\delta T_{21}$ can be written as follows~\cite{Cumberbatch:2008rh,Ciardi:2003hg,prd-edges} 

\beqa
\delta T_{21} =&&16(1-x_e)\left(\frac{\Omega_{b}h}{0.02}\right)\left(\frac{1+z}{10}\frac{0.3}{\Omega_{m}}\right)^{\frac{1}{2}} \nonumber \\
&&\times \left(1-\frac{T_{\rm CMB}}{T_s}\right)\rm mK,
\label{eq:t21}
\eeqa
where $\Omega_{b}$ and $\Omega_{m}$ are the density parameters of baryonic matter and DM, respectively. $h$ is the reduced Hubble constant. $T_s$ is the spin temperature, which is mainly effected by background photons, 
collisions between the particles and resonant scattering of $\rm Ly\alpha$ photons 
(Wouthuysen-Field effect)~\cite{Pritchard:2011xb,Furlanetto:2006jb}. 
Taking into account these factors and with CMB as main background, the spin temperature can be written as follows~\cite{binyue,Cumberbatch:2008rh}

\beqa
T_{s} = \frac{T_{\rm CMB}+(y_{\alpha}+y_{c})T_{k}}{1+y_{\alpha}+y_{c}},
\label{eq:ts}
\eeqa
where $y_{\alpha}$ is related to the Wouthuysen-Field effect and we adopt the formula used in, e.g., Refs.~\cite{binyue,mnras,Kuhlen:2005cm}: 

\beqa
y_{\alpha} = \frac{P_{10}}{A_{10}}\frac{T_{\star}}{T_{k}}{\rm exp}\left[-\frac{0.3(1+z)^{\frac{1}{2}}}{T_{k}^{\frac{2}{3}}\left(1+\frac{0.4}{T_{k}}\right)}\right],
\eeqa
where $A_{10}=2.85\times 10^{-15}s^{-1}$ is the Einstein coefficient of hyperfine spontaneous transition. 
$T_{\star}=0.068\rm K$ corresponds to the energy changes between triplet and singlet states of neutral hydrogen atom. 
$P_{10}$ is the radiative deexcitation rate due to Ly$\alpha$ photons
~\cite{Pritchard:2011xb,Furlanetto:2006jb}. The factor $y_c$ involves collisions between hydrogen atoms and other particles~\cite{binyue,prd-edges,Kuhlen:2005cm,Liszt:2001kh,epjplus-2}, 

\beqa
y_{c} = \frac{(C_{\rm HH}+C_{\rm eH}+C_{\rm pH}){T_{\star}}}{A_{10}T_{k}},
\eeqa  
where $C_{\rm HH, eH, pH}$ are the deexcitation rate due to collisions and the fitted formulas can be found in Refs.~\cite{prd-edges,epjplus-2,Kuhlen:2005cm,Liszt:2001kh}.

In order to explore the conservative allowed space of relevant parameter, following previous works~\cite{DAmico:2018sxd,yinzhema}, 
here we have not included any astrophysical heating source. The main astrophysical source affecting the global 21-cm signal is the Ly$\alpha$ photons related to the Wouthuysen-Field effect~\cite{Furlanetto:2006jb,21cmsolver,yinzhema,Mittal:2020kjs,Gessey-Jones:2022njt,Reis:2021nqf,Mirocha:2020qto,Monsalve:2019baw}. 
Here we have considered that the Ly$\alpha$ photons are mainly from Pop II stars. We take the virial temperature of a 
halo $T_{\rm vir}=10^{4}~\rm K$  corresponding to the minimum halo mass. For the star formation efficiency $f_{\star}$, we have set 
$f_{\star}=0.05$ for our calculations, and we found that larger $f_{\star}$ will slightly increase the amplitude of $\delta T_{21}$ 
at redshift $z=17$. We take the total number of photons from the Pop II stars between the Ly$\alpha$ and the Lyman limits as 
$N_{\rm tot}=9690$. Based on these choices, we can obtain the global 21-cm signal at redshift $z=17$ with the maximum amplitude 
allowed within a reasonable range of parameters.

\begin{figure}[htp]
\centering
\includegraphics[width=0.85\linewidth]{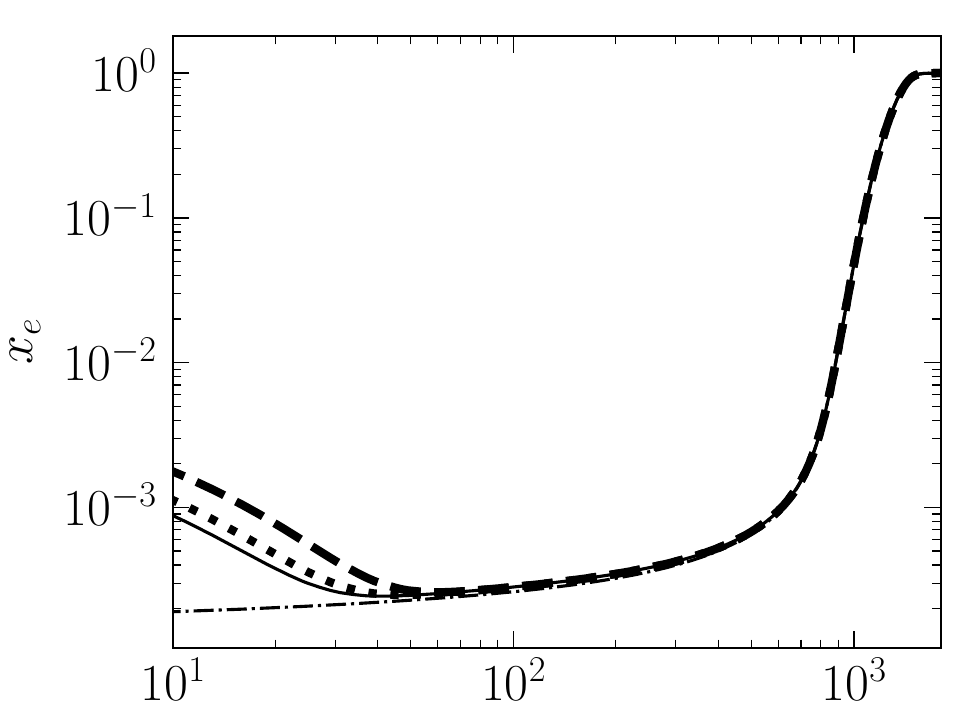}
\hfill
\includegraphics[width=0.85\linewidth]{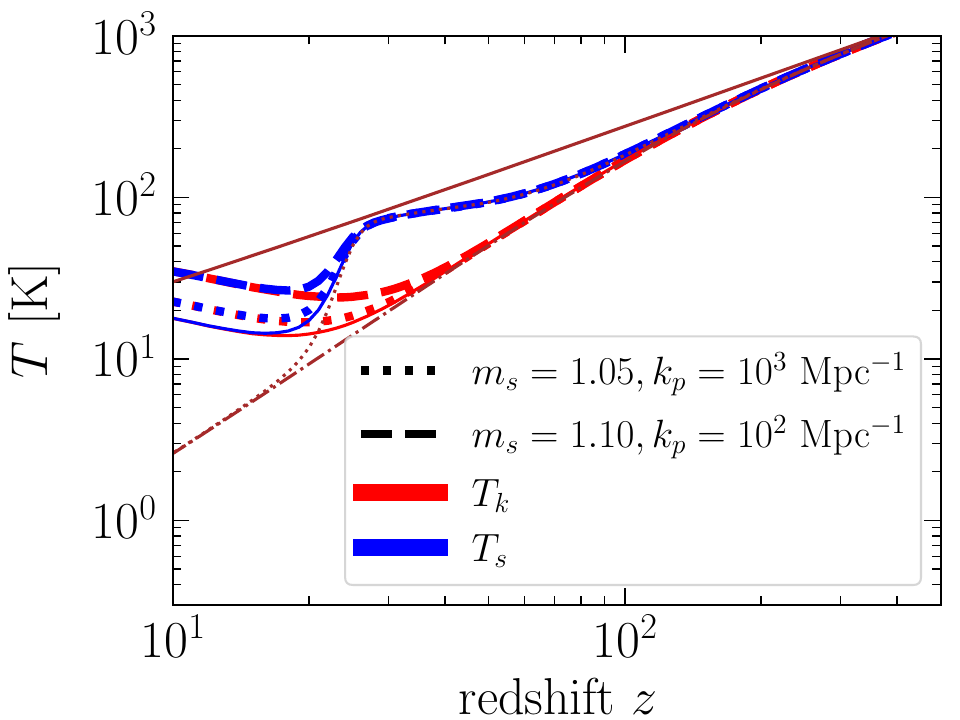}
\caption{The evolution of $x_e$, $T_k$ and $T_s$ with redshift $z$ for different values of $m_s$ and $k_p$. Here we have set $m_{\chi} = 100~\rm GeV$ and $\left<\sigma v\right> = 3\times 10^{-26}~\rm cm^{-3}s^{-1}$. 
We also plot the standard scenario for comparison ($m_{s}=n_s=0.96$, thin solid lines). The case without standard astrophysical heating sources 
is also shown ($x_e$: thin dot-dashed black line, $T_k$: thin dot-dashed brown line, $T_s$: thin dotted brown line). 
The temperature of CMB is also shown in the thin solid brown line.}
\label{fig:xe-tk-ts}
\end{figure}

After deriving the energy release rate per unit volume due to DM annihilation as shown in Eq.~(\ref{eq:de_dvdt}), 
one can get the changes of $x_e$, $T_k$ and $T_s$ with redshift $z$ using Eqs.~(\ref{eq:xe}), (\ref{eq:tk}) and (\ref{eq:ts}). 
In order to include the effects of DM annihilation, we have modified the public code RECFAST 
in CAMB\footnote{https://camb.info/} to solve the differential equations numerically~\cite{mnras,DM_2015,xlc_decay,lz_decay,prd-2020,yinzhema}.
Then the differential brightness temperature $\delta T_{21}$ can be obtained with Eq.~(\ref{eq:t21}). 
In Fig.~\ref{fig:xe-tk-ts}, we plot the evolution of $x_e$, $T_k$ and $T_s$ for different values of $m_s$ and $k_p$. 
Compared with the standard scenario ($m_{s}=n_s=0.96$, thin solid lines), the ionization fraction, kinetic temperature and spin temperature 
are all increased. Here we have set the canonical value of DM annihilation cross section as $\left<\sigma v\right> = 3\times 10^{-26}~\rm cm^{-3}s^{-1}$ and $b\bar b$ channel for our calculations. 
 
\begin{figure}[htp]
\centering
\includegraphics[width=0.85\linewidth]{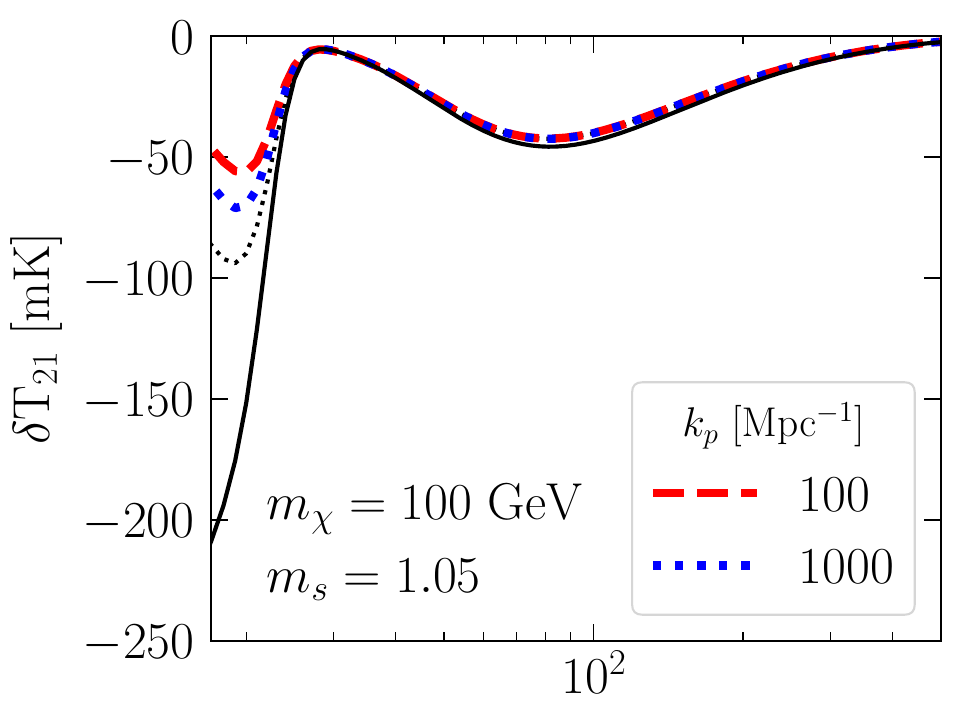}
\hfill
\includegraphics[width=0.85\linewidth]{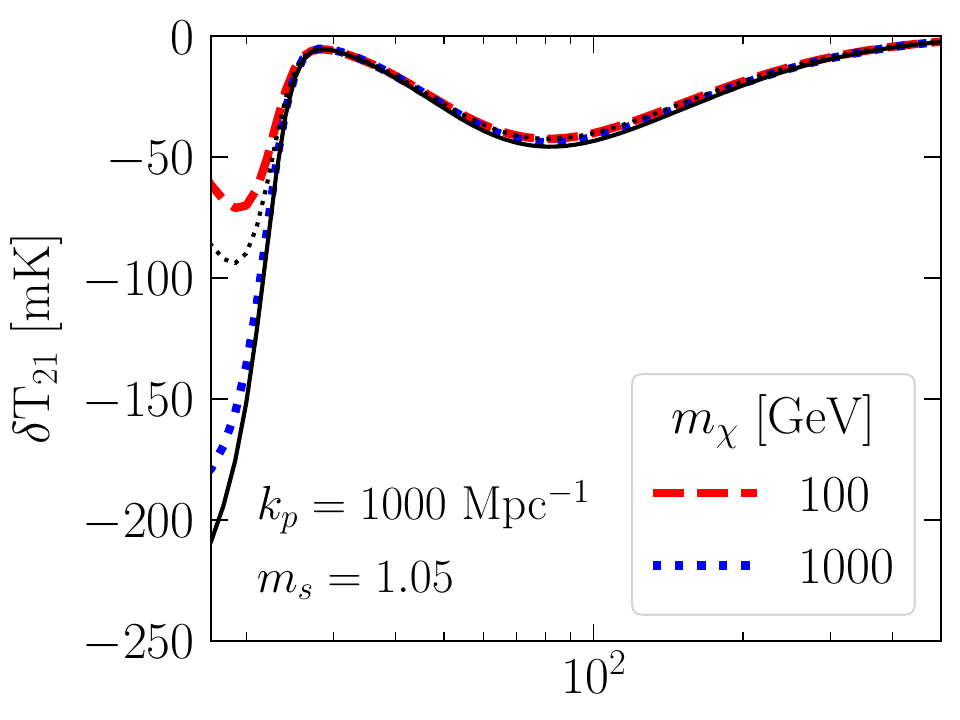}
\hfill
\includegraphics[width=0.85\linewidth]{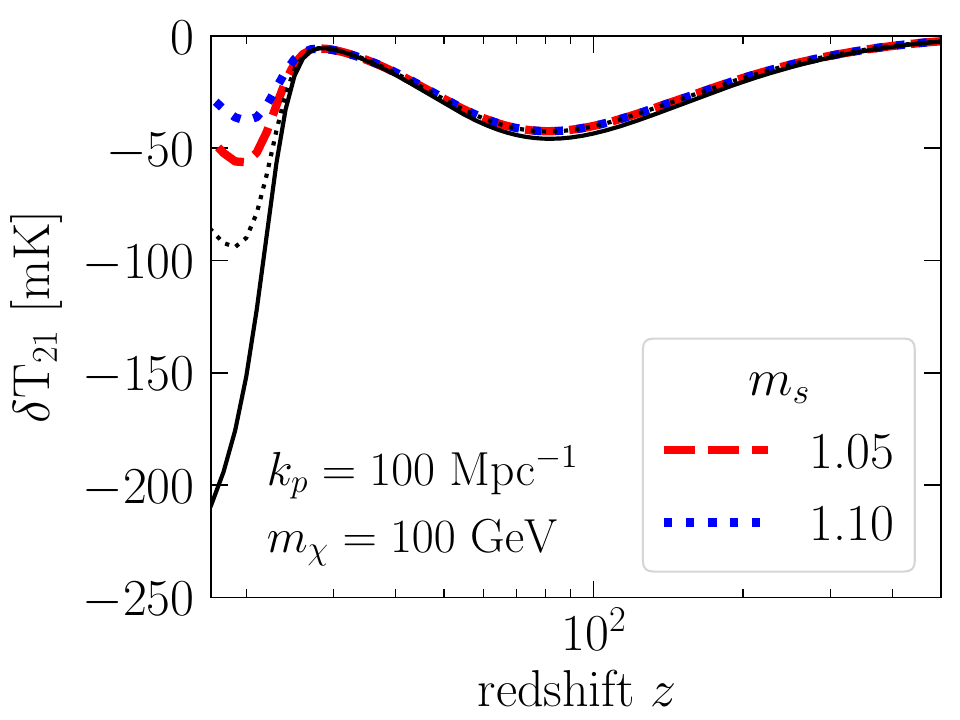}
\caption{The evolution of differential brightness temperature $\delta T_{21}$ with redshift $z$ for different values of $k_p$, $m_{\chi}$ and $m_s$. We have set $\left<\sigma v\right> = 3\times 10^{-26}~\rm cm^{-3}s^{-1}$ and $b\bar b$ channel for our calculations. For comparison, we also show the case without any heating sources (thin solid black line) and the case for the standard matter power spectrum $m_{s}=n_{s}$ with DM annihilation ($m_{\chi}=100~\rm GeV$, thin dotted black line).}
\label{fig:t21_com}
\end{figure}

In Fig.~\ref{fig:t21_com}, the evolution of $\delta T_{21}$ for different values of $k_p$, 
$m_{\chi}$ and $m_s$ are shown respectively. For comparison, we also plot the default case with no DM annihilation (thin solid black line) 
and the standard case with no deviation of the matter power spectrum at small scales (thin dotted black line). 
For fixed DM mass $m_{\chi}$ and power law index of the deviation $m_s$ (top panel in Fig.~\ref{fig:t21_com}), 
smaller pivot scale results in an 
increase of the number density of DM halos at small masses. Therefore, 
the absorption amplitude of the global 21-cm signal is reduced compared with the standard scenario. 
For fixed pivot scale $k_p$ and power law index of the deviation $m_s$ (middle panel in Fig.~\ref{fig:t21_com}), 
lighter DM corresponds to a larger DM number density. Since the DM annihilation rate is proportional to 
the squared number density, much more energy is injected into the IGM, causing a reduction of the absorption amplitude 
of the global 21-cm signal. For fixed pivot scale $k_p$ and DM mass $m_{\chi}$ 
(bottom panel in Fig.~\ref{fig:t21_com}), larger power low index 
of the deviation $m_s$ also results in an increase of the number density of DM halos at small masses. Similar to 
the case of changing pivot scale, the absorption amplitude of the global 21-cm signal is decreased for 
larger $m_s$ compared with the scenario of no deviation.

%%%%%%%%%%%%%%%%%%%%%%%%%%%%%%%%%%Figure 2 begin %%%%%%%%%%%%%%%%%%%%%%%%%%%%%%%%%%%%%%

%%%%%%%%%%%%%%%%%%%%%%%%%%%%%%%%%Figure 2 end %%%%%%%%%%%%%%%%%%%%%%%%%%%%%%%%%%%%%%%%%

%%%%%%%%%%%%%%%%%%%%%%%%%%% Figure 3 begin  %%%%%%%%%%%%%%%%%%%%%%%%%%%%%%%

%%%%%%%%%%%%%%%%%%%%%%%%%% Figure 3 end %%%%%%%%%%%%%%%%%%%%%%%%%%%%%%%%%

\begin{figure}[htp]
\centering
\includegraphics[width=0.95\linewidth]{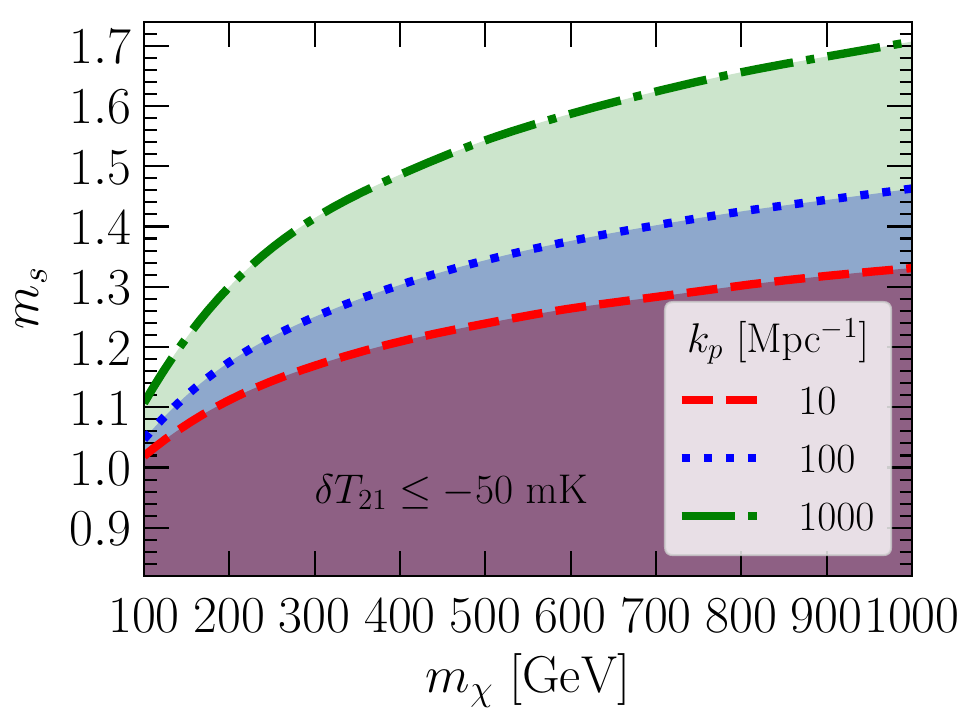}
\caption{The allowed space of parameter 
$m_s$ (shaded areas) for different pivot scales $k_p=10, 100~{\rm and}~1000~\rm Mpc^{-1}$ by requiring 
the differential brightness temperature $\delta T_{21}\le -50~\rm mK$. 
We have set $\left<\sigma v\right> = 3\times 10^{-26}~\rm cm^{-3}s^{-1}$ and 
$b\bar b$ channel for our calculations. }
\label{fig:ms_cons}
\end{figure}

In view of the results of the EDGES experiment, by requiring 
the differential brightness temperature $\delta T_{21} \le -50~\rm mK$ at redshift $z=17$, 
we explore the allowed space of parameter $m_s$ for different pivot scales 
$k_{p}=10,100~{\rm and}~1000~\rm Mpc^{-1}$, which is shown 
in Fig.~\ref{fig:ms_cons} . 
From this plot, it can be seen that smaller DM mass or pivot scale corresponds 
to a smaller value of $m_s$. In Ref.~\cite{03480}, the authors derived the upper limits 
on $m_s$ using the CMB observations. They found that for the parameter $f\left<\sigma v\right>/m_{\chi} 
= 3\times 10^{-28}~\rm cm^{3}~s^{-1}~Gev^{-1}$, the upper limit is $m_{s}=1.43(1.63)$ 
for $k_{p}=100(1000)h~\rm Mpc^{-1}$, and it is roughly weaker than our result for 
the DM mass range considered here.

%%%%%%%%%%%%%%%%%%%%%%%%%%%%% Figure 4 begin %%%%%%%%%%%%%%%%%%%%%%%%%%%

%%%%%%%%%%%%%%%%%%%%%%%%%%%%% Figure 4 end %%%%%%%%%%%%%%%%%%%%%%%%%%%

Note that here we have not included the standard astrophysical heating sources, e.g., the X-ray from stars, which can also heat the IGM and result in the increase of $x_e$, $T_k$ and $T_s$ in lower redshifts~\cite{Pritchard:2011xb,Furlanetto:2006jb,binyue}. For this case, compared with our results, the amplitude of differential brightness temperature $\delta T_{21}$ at redshift $z=17$ will become smaller, 
resulting in a lower allowed value of $m_s$.

In this work, we have used the canonical value of DM annihilation cross section for our calculations. Many astronomical observations have been used to constrain $\left<\sigma v\right>$ depending on the DM mass~\cite{Slatyer:2015jla,Kawasaki:2021etm,Li:2013qya,DiMauro:2021qcf,Fermi-LAT:2013sme}. The authors of~\cite{Kawasaki:2021etm}, for instance, have used the Planck-2018 datasets to get the constraints and found $\left<\sigma v\right> \lesssim 3\times 10^{-25}(10^{-24})~\rm cm^{-3}s^{-1}$ for $m_{\chi}=100(1000)~\rm GeV$. As shown in Eq.~(\ref{eq:de_dvdt}), larger value of $\left<\sigma v\right>$ will result in the larger energy release rate per unit volume, 
and it is excepted that the final allowed value of $m_s$ will become smaller.

Note that the final allowed space of parameter $m_s$ can be effected by relevant parameters, and these parameters would be degenerate with each other. 
A complete way to deal with this issue is combining the observed data of the EDGES to obtain the distribution and correlation of parameters by using the MCMC method. We will address this issue in future work. 
The similar effects can be also caused by the primordial black holes.\footnote{Y.Yang et al. in preparation.}

\section{conclusions}
In the standard scenario, the matter power spectrum has a form of $P_{m}(k) \sim k^{n_s}$. 
Many relevant theories indicate that the matter power spectrum could be deviated at small scales 
while being consistent with the available astronomical observations. In this work, we have investigated 
the impact of this kind of deviation on the global 21-cm signal in the cosmic dawn, taking into account DM annihilation. 
Specifically, we have adopted a power law growth of the matter power spectrum at small scales, 
$P_{m} \sim k_{p}^{n_s}(k/k_{p})^{m_s}$ for $k>k_p\simeq 10~\rm Mpc^{-1}$. The deviation of the matter power spectrum 
at small scales results in an increase of the comoving number density of DM halos at small masses. 
The energy release rate per unit volume due to DM annihilation becomes larger compared with 
the standard scenario, resulting in the changes of the thermal history of IGM and then 
the evolution of the global 21-cm signal. 
The absorption amplitude of the global 21-cm signal is reduced for smaller pivot scale $k_p$ 
or larger power law index $m_s$. Smaller DM mass $m_{\chi}$ can also decrease the absorption amplitude of 
the global 21-cm signal due to the larger annihilation rate. In view of the results of the EDGES experiment, 
we have explored the allowed parameter space of the power law index $m_s$ for different pivot scales 
by requiring the differential brightness temperature $\delta T_{21} \le -50~\rm mK$. 
Smaller DM mass or pivot scale results in a lower allowed value of $m_s$. 
For a DM mass, e.g., $m_{\chi}=100(1000)~\rm GeV$, the largest allowed value is $m_s=1.05(1.46)$ 
for the pivot scale $k_{p} = 100~\rm Mpc^{-1}$. 

Note that we have considered the global 21-cm signal in the cosmic dawn that 
can be influenced by many other astrophyscial factors. 
The global 21-cm signal in the dark ages ($30\lesssim z\lesssim 300$) can also be effected 
by the deviation of the matter power spectrum at small scales. Compared with the standard scenario, 
the global 21-cm signal in the dark ages is very little influenced by the astrophysical processes. 
Therefore, it is expected that the future detection of the global 21-cm signal 
(or the 21-cm power spectrum) 
in the dark ages by, e.g., the radio telescopes on the moon or satellites around 
a low lunar orbit~\cite{Burns:2019zia,2017arXiv170200286P,2019arXiv190710853C,Burns:2021ndk,Burns_2020,Burns:2021pkx}, 
could give better constraints on the deviation of the matter power spectrum at small scales.

\section{Acknowledgements}
The authors would like to thank the anonymous referees for their very helpful
comments and suggestions. Y. Yang thank Bin Yue for very useful discussions. 
Y. Yang is supported by the Shandong Provincial Natural Science Foundation (Grant No. ZR2021MA021). 
X. Li is supported by the Youth Innovations and Talents Project of Shandong Provincial Colleges 
and Universities (Grant No. 201909118). G. Li is supported by the Taishan Scholar Project of Shandong Province 
(Grant No. tsqn202103062).
\

\bibliographystyle{apsrev4-title}
\bibliography{matter_power_spectrum_at_small_scale_third}
%\bibliography{refs}
  
\end{document}